\documentclass[preprint2]{emulateapj}
\usepackage{graphicx}
\usepackage{txfonts}
\usepackage{url}

\newcommand{\EQ}{\begin{equation}}
\newcommand{\EN}{\end{equation}}
\newcommand{\EQA}{\begin{eqnarray}}
\newcommand{\ENA}{\end{eqnarray}}

\newcommand{\Eq}[1]{equation~(\ref{#1})}

\newcommand{\Sec}[1]{\S\ref{#1}}

\newcommand{\Fig}[1]{Fig.~\ref{#1}}
\newcommand{\FFig}[1]{Figure~\ref{#1}}
\newcommand{\Tab}[1]{Table~\ref{#1}}

\newcommand{\bra}[1]{\langle #1\rangle}

{}
{}

{}
{}
{}
{}
{}
{}
{}
{}
{}
\newcommand{\meanBB}{\overline{\mbox{\boldmath $B$}}{}}{}
{}
{}
{}
{}
{}

{}
{}
{}
{}
{}

\newcommand{\UU}{\mbox{\boldmath $U$} {}}

\newcommand{\BB}{\mbox{\boldmath $B$} {}}

\newcommand{\JJ}{\mbox{\boldmath $J$} {}}

\newcommand{\AAA}{\mbox{\boldmath $A$} {}}

\newcommand{\ff}{\mbox{\boldmath $f$} {}}

\newcommand{\WW}{\mbox{\boldmath $W$} {}}

\newcommand{\nab}{\mbox{\boldmath $\nabla$} {}}

\newcommand{\SSSS}{\mbox{\boldmath ${\sf S}$} {}}

\newcommand{\dd}{{\rm d} {}}
\newcommand{\const}{{\rm const}  {}}

\def\la{\mathrel{\mathchoice {\vcenter{\offinterlineskip\halign{\hfil
$\displaystyle##$\hfil\cr<\cr\sim\cr}}}
{\vcenter{\offinterlineskip\halign{\hfil$\textstyle##$\hfil\cr<\cr\sim\cr}}}
{\vcenter{\offinterlineskip\halign{\hfil$\scriptstyle##$\hfil\cr<\cr\sim\cr}}}
{\vcenter{\offinterlineskip\halign{\hfil$\scriptscriptstyle##$\hfil\cr<\cr\sim\cr}}}}}

\def\Pm{\mbox{\rm Pr}_M}
\def\Rm{\mbox{\rm Re}_M}

\def\Rey{\mbox{\rm Re}}

\def\cs{c_{\it s}}
\def\kf{k_{\it f}}
\def\urms{u_{\rm rms}}
\def\Brms{B_{\rm rms}}
\def\etat{\eta_{\it t}}

\def\half{{\textstyle{1\over2}}}

\def\onethird{{\textstyle{1\over3}}}

\newcommand{\K}{\,{\rm K}}
\newcommand{\g}{\,{\rm g}}

\newcommand{\cm}{\,{\rm cm}}

\newcommand{\yapj}[3]{ #1, {ApJ,} {#2}, #3}

\newcommand{\yapjl}[3]{ #1, {ApJ,} {#2}, #3}

\newcommand{\yana}[3]{ #1, {A\&A,} {#2}, #3}

\newcommand{\yjfm}[3]{ #1, {J.\ Fluid Mech.,} {#2}, #3}

\newcommand{\ypf}[3]{ #1, {Phys.\ Fluids,} {#2}, #3}

\newcommand{\yjetp}[3]{ #1, {Sov.\ Phys.\ JETP,} {#2}, #3}

\newcommand{\yprl}[3]{ #1, {Phys.\ Rev.\ Lett.,} {#2}, #3}

\newcommand{\ymn}[3]{ #1, {MNRAS,} {#2}, #3}

\newcommand{\ypre}[3]{ #1, {Phys.\ Rev.\ E,} {#2}, #3}

\newcommand{\yjour}[4]{ #1, {#2}, {#3}, #4}

\newcommand{\ybook}[3]{ #1, {#2} (#3)}

\begin{document}

\title{Large-scale dynamos at low magnetic Prandtl numbers}
\author{Axel Brandenburg}

\affil{
NORDITA, AlbaNova University Center,
Roslagstullsbacken 23, SE-10691 Stockholm, Sweden
}

\email{brandenb@nordita.org
 ($ $Revision: 1.57 $ $)
}

\begin{abstract}
Using direct simulations of hydromagnetic turbulence driven by random
polarized waves it is shown that dynamo action is possible over a wide
range of magnetic Prandtl numbers from $10^{-3}$ to 1.
Triply periodic boundary conditions are being used.
In the final saturated state the resulting magnetic field has a
large-scale component of Beltrami type.
For the kinematic phase, growth rates have been determined for magnetic
Prandtl numbers between 0.01 and 1, but only the case with the smallest
magnetic Prandtl number shows large-scale magnetic fields.
It is less organized than in the nonlinear stage.
For small magnetic Prandtl numbers the growth rates are comparable
to those calculated from an alpha squared mean-field dynamo.
In the linear regime the magnetic helicity spectrum has a short inertial
range compatible with a $-5/3$ power law, while in the nonlinear regime
it is the current helicity whose spectrum may be compatible with such a law.
In the saturated case, the spectral magnetic energy in the inertial range
is in slight excess over the spectral kinetic energy, although for
small magnetic Prandtl numbers the magnetic energy spectrum
reaches its resistive cut off wavenumber more quickly.
The viscous energy dissipation declines with the square root of the
magnetic Prandtl number, which implies that most of the energy is
dissipated via Joule heat.
\end{abstract}


\section{Introduction}

Many astrophysical plasmas are turbulent and tend to be magnetized.
The magnetic fields can have typical length scales that are either larger
or smaller than that of the energy-carrying eddies.
We speak then correspondingly of large-scale or small-scale dynamos.
Small-scale dynamos can already work in statistically mirror-symmetric
isotropic homogeneous turbulence, whereas large-scale dynamos require
in general a departure from parity-invariant or mirror-symmetric flows.
The excitation conditions of small-scale dynamos depend sensitively
on the value of the magnetic Prandtl number, i.e.\
the ratio of kinematic viscosity to magnetic diffusivity, $\Pm=\nu/\eta$,
where $\nu$ is the kinematic viscosity and $\eta$ the magnetic diffusivity.
This sensitivity is related to the fact that in the {\it kinematic} regime
the spectral magnetic energy is peaked at the resistive scale.
As was pointed out originally by Rogachevskii \& Kleeorin (1997),
and more recently by Boldyrev \& Cattaneo (2004), the slope of the
kinetic energy spectrum is important for the
onset of small-scale dynamo action.
It matters therefore whether the resistive scale lies in the
viscous range ($\Pm\approx1$), within the inertial range ($\Pm\la0.1$),
or right within the range where the bottleneck occurs ($\Pm\approx0.1$).
The bottleneck effect refers to the spectral subrange just before the
dissipation range where the kinetic energy spectrum is shallower than
in the inertial range.
Within the bottleneck range the velocity increments diverge even more
strongly with decreasing separation than in the inertial range, making
dynamo action harder still.
Indeed, for small values of $\Pm$ the critical value of the magnetic Reynolds
number above which small-scale dynamo action occurs increases therefore
sharply toward $\Pm=0.1$ (Schekochihin et al.\ 2005), and then decreases
slightly for $\Pm<0.05$ (Iskakov et al.\ 2007).
However, even with the computing power available today, direct simulations
of small-scale dynamo action are still only marginally possible at such
small values of $\Pm$.

For certain types of flows dynamo action is easier to achieve
even though $\Pm$ is small.
The Taylor-Green flow is an example where the critical value of the
magnetic Reynolds number becomes constant for $\Pm<0.1$
(Ponty et al.\ 2004, 2005).
In this flow there can be large-scale patches with finite kinetic
helicity of opposite sign.
A completely different example is fully helical turbulence where the excitation
condition for dynamo action is virtually unchanged as $\Pm$ decreases
from 1 to 0.1 (Brandenburg 2001, hereafter B01).
For an ABC-flow dynamo, Mininni (2007) found a weak dependence
of the threshold value of the magnetic Reynolds number $\Rm$ on $\Pm$.
For $\Pm<0.1$ the threshold value seemed to become asymptotically
independent of $\Pm$ and dynamo action was demonstrated for values
of $\Pm$ down to $5\times10^{-3}$.

In many astrophysical bodies, $\Pm$ is indeed rather small (around $10^{-5}$).
Such systems still possess dynamo action and can have large-scale
magnetic fields.
It is likely that such systems belong to the second class of systems
where the excitation conditions are not drastically altered toward
small values of $\Pm$.
Indeed, large-scale magnetic fields are found regardless of whether
$\Pm$ is small (e.g., the Sun and other stars with outer convection zones,
as well as planets) or large (e.g., in spiral galaxies, because of their
low densities).

The purpose of this paper is to point out that a strong $\Pm$
dependence does not occur in systems where the magnetic field generation
is predominantly due to a large-scale dynamo.
Such systems have been studied in idealized settings such as
periodic boxes using explicit forcing functions for driving the turbulence.
This has significant advantages in that periodic boundary conditions
can be used, energy spectra are easily computed and, most importantly,
isotropy and homogeneity eases comparison with turbulence theory.
A disadvantage is that the magnetic helicity can only change on resistive
timescales, which slows down the saturation (B01).

With these provisions in mind, we consider now simulations of maximally
helical turbulence in triply periodic boxes where we keep in most cases
the fluid Reynolds number, $\Rey=\urms/\nu\kf$, constant
and vary the magnetic Reynolds number, $\Rm=\urms/\eta\kf$, and thereby
$\Pm$ ($\equiv\Rm/\Rey$).
Here, $\urms$ is the rms velocity of the turbulence and $\kf$ is the forcing
wavenumber.
According to B01 the dynamo should be excited
whenever the domain is large enough (2--3 times larger than the forcing scale)
and the magnetic Reynolds number exceeds unity ($\Rm\geq1.1...1.4$ or so).
This was confirmed for magnetic Prandtl numbers as low as 0.1.
In the present work we consider kinematic dynamo action down to values
of $\Pm=10^{-2}$ and nonlinear saturated dynamos down to $\Pm=10^{-3}$.

\section{The method}

We solve the hydromagnetic equations for velocity $\UU$, logarithmic
density $\ln\rho$, and magnetic vector potential $\AAA$ for an isothermal gas
in the presence of an externally imposed helical forcing function $\ff$,
\EQ
{\partial\UU\over\partial t}=-\UU\cdot\nab\UU-\cs^2\nab\ln\rho
+\ff+\rho^{-1}\Big(\JJ\times\BB+\nab\cdot2\rho\nu\SSSS\Big) \, ,
\label{dUU}
\EN
\EQ
{\partial\ln\rho\over\partial t}=-\UU\cdot\nab\ln\rho-\nab\cdot\UU,
\label{dlnrho}
\EN
\EQ
{\partial\AAA\over\partial t}=\UU\times\BB-\mu_0\eta\JJ.
\label{indEq}
\EN
Here, $\BB=\nab\times\AAA$ is the magnetic field, $\JJ=\nab\times\BB/\mu_0$
is the current density, $\mu_0$ is the vacuum permeability,
$\cs$ is the isothermal speed of sound, and
${\sf S}_{ij}={1\over2}(U_{i,j}+U_{j,i})-{1\over3}\delta_{ij}\nab\cdot\UU$
is the traceless rate of strain tensor.
We consider a triply periodic domain of size $L^3$, so the smallest
wavenumber in the domain is $k_1=2\pi/L$.
The forcing function consists of eigenfunctions of the curl operator
with positive eigenvalues and is therefore fully helical with
$\ff\cdot\nab\times\ff=k\ff^2$,
where $3.5\leq k/k_1\leq4.5$ is wavenumber interval of the forcing function,
whose average value is referred to as $\kf=4\,k_1$.
The amplitude of $\ff$ is such that the Mach number is $\urms/\cs\approx0.1$,
so compressive effects are negligible (Dobler et al.\ 2003).

The initial conditions consist of a weak Beltrami field.
The initial velocity is zero and the initial density is uniform
with $\rho=\rho_0=\const$.
Note that the volume-averaged density remains constant,
i.e., $\bra{\rho}=\rho_0$.

The model is equivalent to that of B01, except that there the value of
$\kf/k_1$ was chosen to be either 5 or 30.
In order for large-scale dynamo action to be possible, $\kf/k_1$ must
at least be larger than 2 (Haugen et al.\ 2004), but 3 is already
sufficient (Brandenburg et al.\ 2008).
In the kinematic regime the fastest growing mode is expected to have
the wavenumber $\kf/2$, so in order that this wavenumber is distinct
from $k_1$, we have chosen $\kf/k_1=4$ throughout this paper.

\section{Results}

We begin by presenting results for $\Rey\approx670$ where we vary $\Pm$
in the range $0.01\leq\Pm\leq1$, i.e., $\Rm$ is varied in the range
$6.7\leq\Rm\leq670$.
The dynamo is excited in all those cases, but the growth rate $\lambda$ varies.
We consider first the kinematic regime where the magnetic field is weak
and turn then to the nonlinear regime where the magnetic field has saturated.
Most of the results presented below have been obtained at a resolution
of $512^3$ meshpoints.
The solution was first
evolved at lower resolution ($128^3$ meshpoints), then remeshed to twice
the resolution, again evolved for some time, and finally remeshed to
$512^3$ meshpoints, and again evolved for some time.
Data for the kinematic regime are only used after the initial transients
have disappeared and a clear exponential growth has developed at all
length scales for at least some 40 turnover times (also for the runs
with a resolution of $512^3$ meshpoints).
The run with $128^3$ meshpoints has been evolved all the way into
saturation, and it was then remeshed twice by a factor of 2, just like
in the kinematic regime.

\begin{figure*}[t!]
\begin{center}\plotone{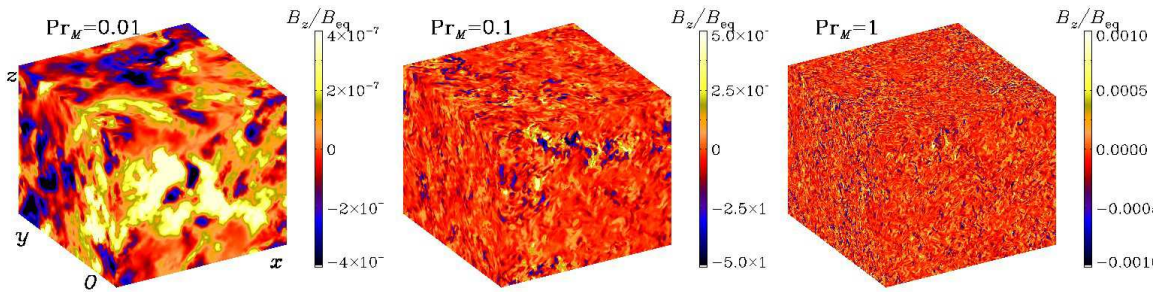}\end{center}\caption[]{
Visualization of $B_z$ for $\Pm=0.01$, 0.1, and 1 at $\Rey=670$.
Note the emergence of a large-scale pattern for $\Pm=0.01$.
For $\Pm=0.1$ there are only a few extended patches and for $\Pm=1$
the field is completely random and of small scale only.
The orientation of the axes is indicated for the first panel,
and is the same for all other panels.
}\label{B}\end{figure*}

\subsection{Field structure in the kinematic regime}

Visualizations of one component of the magnetic field show the emergence
of a large-scale magnetic field for small values of $\Pm$.
This is clearly demonstrated in \Fig{B}, where we see for $\Pm=0.01$ a
large-scale pattern with a systematic variation in the $y$ direction.
For $\Pm=0.1$ there is no such variation, although there are some extended
patches in which the field orientation is the same.
For $\Pm=1$ even this is no longer the case and the field appears
completely random with small-scale variations only.

We emphasize that random and patch-like structures only occur in the
kinematic regime.
In the saturated regime a large-scale field emerges in all cases.
This will be discussed in \Sec{VelocityPattern}.

\subsection{Growth rates}

\begin{figure}[t!]
\begin{center}\plotone{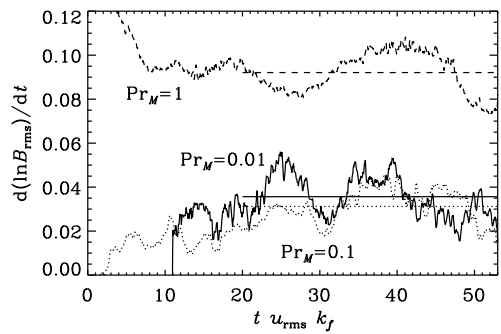}\end{center}\caption[]{
Instantaneous growth rate for runs with different values
of $\Pm$ and $512^3$ meshpoints.
The straight lines give the growth rates obtained by averaging over
the indicated time interval.
The solid, dotted, and dashed lines are for $\Pm=0.01$, 0.1, and 1, respectively.
}\label{pcomp}\end{figure}

\begin{figure}[t!]
\begin{center}\plotone{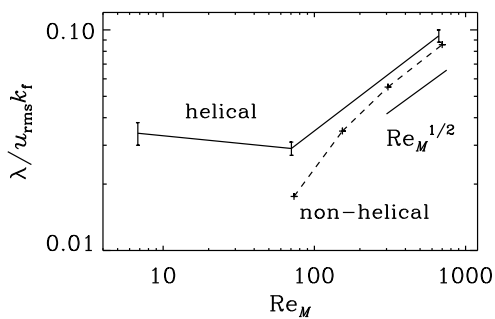}\end{center}\caption[]{
Dynamo growth rates of the rms magnetic field for
helical turbulence with $\Rey=670$ (solid line)
compared with growth rates for non-helical turbulence (dashed lines;
adapted from Haugen et al.\ 2004).
}\label{pgrowth_haugen_etal}\end{figure}

The growth rate is calculated as the average of the instantaneous growth
rate, $\dd\ln\Brms/\dd t$.
Examples are shown in \Fig{pcomp} for runs with $\Rey=670$ and different
values of $\Pm$ using $512^3$ meshpoints.
In \Fig{pgrowth_haugen_etal}, we show growth rates
normalized by $\urms\kf$ (inverse turnover times) as a function of $\Rm$ for
three values of $\Rm$ and compare with the corresponding results for
non-helical turbulence forced at larger scales in the wavenumber interval
$1\leq k/k_1\leq2$ (Haugen et al.\ 2004).
In that case we use $\kf=1.5$ for the average value.
The small-scale dynamo is then only excited when $\Rm\ge35$.
For $\Rm\ge100$ the growth rates for helical turbulence with $\kf=4$
are quite similar to those of non-helical turbulence with $\kf=1.5$.
We have also calculated growth rates for the non-helical case with
$\kf=4$ and find the same values as in the helical case.
We note that in all cases, and even for small values of $\Pm$, the
growth rates based on the rms magnetic field are equal to those based
on the rms values of the mean fields obtained by averaging over
any two coordinate directions.

In both helical and non-helical cases,
when $\Rm$ is large enough, $\lambda$ increases like
$\Rm^{1/2}$, as expected (Schekochihin et al.\ 2004).
This is because the eddy turnover rate at the resistive scale is
$\propto k_\eta^{2/3}$, but because $k_\eta/\kf\propto\Rm^{3/4}$
we have $\lambda\propto\Rm^{1/2}$.
However, when there is also large-scale dynamo action,
one expects there to be a lower bound for
$\lambda$ given by the growth rate for the large-scale dynamo,
$\lambda_{\rm LS}$.
Using the theory for an $\alpha^2$ dynamo (Moffatt 1978,
Krause \& R\"adler 1980), we have
\EQ
\lambda_{\rm LS}=|\alpha k|-(\eta+\etat)k^2,
\label{lambdaLS}
\EN
where $\alpha$ is a pseudo scalar (the $\alpha$ effect) and $\etat$ is
the turbulent magnetic diffusivity.
For fully helical turbulence we have $|\alpha|\approx\urms/3$ and
$\etat\approx\urms/3\kf$ (Sur et al.\ 2008), so we can write the growth rate as
\EQ
{\lambda_{\rm LS}\over\urms\kf}=\onethird{k_1\over\kf}
\left[1-{k_1\over\kf}\left(1+3\Rm^{-1}\right)\right],
\EN
where we have put $|k|=k_1$.
Over the parameter range considered in this paper ($\Rm\ge6.7$ and $\kf/k_1=4$),
$\lambda_{\rm LS}/\urms\kf$ increases only slightly from 0.053 to 0.062
as $\Rm$ increases.

The result shown in \Fig{pgrowth_haugen_etal} gives values that are
systematically below $\lambda_{\rm LS}$.
There could be two reasons for this discrepancy.
On the one hand, the accuracy of the estimates 
$|\alpha|\approx\urms/3$ and $\etat\approx\urms/3\kf$
may not be good enough.
On the other hand, \Eq{lambdaLS} is only an approximation in cases
where $\lambda_{\rm LS}\neq0$, because then memory effects become
important.
This means that, when allowing $\alpha$ and $\etat$ to be integral
kernels in time, they are no longer proportional to $\delta$ functions,
but have finite widths in time.
This effect has recently been studied by Hubbard \& Brandenburg (2008)
and can be quite dramatic in some cases.

\begin{figure}[t!]
\begin{center}\plotone{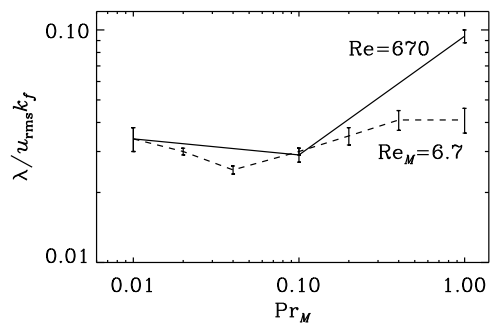}\end{center}\caption[]{
Dependence of dynamo growth rates of the rms magnetic field
on $\Pm$ for helical turbulence with
$\Rm=6.7$ (dashed line) and $\Rey=670$ (solid line).
Here the solid line corresponds to the  solid line in
\Fig{pgrowth_haugen_etal}.
}\label{pgrowth}\end{figure}

In \Fig{pgrowth_haugen_etal}, we have varied $\Rm$ by changing $\Pm$
and keeping $\Rey=\const=670$.
We can therefore also consider this graph as a representation of the
magnetic Prandtl number dependence of $\lambda$.
However, $\Pm$ can also be changed while keeping $\Rm=\const=6.7$.
The corresponding result is shown in \Fig{pgrowth} and compared with the
previous case.
The two graphs are in reasonable agreement for small values of $\Pm$,
but the rise of $\lambda$ for $\Rey=670$ around $\Pm=1$ is not seen
in the case with $\Rm=6.7$.
This suggests that the transition from a purely large-scale turbulent dynamo
to a mixed large-scale and small-scale turbulent dynamo requires values of
$\Rm$ above some critical value (somewhere between 10 and 100),
and is not just determined by the value of $\Pm$.

\subsection{Spectra}
\label{Spectra}

The transition from a purely large-scale turbulent dynamo to a mixed
large-scale and small-scale turbulent dynamo is accompanied by
characteristic changes in the spectral properties of the magnetic field.
In the following we employ shell-integrated spectra of kinetic and
magnetic energy, $E(k)$ and $M(k)$, respectively, as well as of
kinetic and magnetic helicities, $F(k)$ and $H(k)$, respectively.
These spectra are normalized such that
$\int E(k)\,\dd k=\half\bra{\UU^2}\equiv E$,
$\int M(k)\,\dd k=\half\bra{\BB^2}\equiv M$,
$\int F(k)\,\dd k=\bra{\WW\cdot\UU}$, and
$\int H(k)\,\dd k=\bra{\AAA\cdot\BB}$, where
$\WW=\nab\times\UU$ is the vorticity.

\begin{figure}[t!]
\begin{center}\plotone{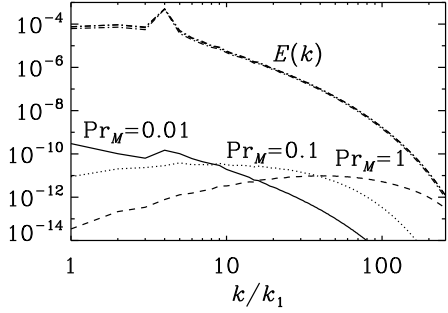}\end{center}\caption[]{
Spectra of kinetic and magnetic energies in the kinematic regime
for $\Pm=0.01$, 0.1, and 1.
}\label{pspec_aver_comp_lopm}\end{figure}

In \Fig{pspec_aver_comp_lopm}, we plot $E(k)$ and $M(k)$
for three cases with $\Pm=1$, 0.1, and 0.01, keeping $\Rey=670$ in all cases.
The magnetic energy spectra are compensated by $\exp(-\lambda t)$,
where $\lambda$ is the numerically determined growth rate for each run,
and then averaged in time.
We consider here only the kinematic regime when the magnetic energy is weak.
The kinetic energy spectra are then always the same.
Since the magnetic energy is weak, we have scaled the magnetic energy spectra
for different $\Pm$ to an arbitrarily chosen reference value of $10^{-6}$
below the kinetic energy spectrum.
For $\Pm=1$ the magnetic energy seems to follows an approximate Kazantsev (1968) spectrum
with a range proportional to $k^{3/2}$ and is peaked at the resistive scale
near $k/k_1=50$.
For smaller values of $\Pm$ the peak of magnetic energy moves to smaller
wavenumbers.

\begin{figure}[t!]
\begin{center}\plotone{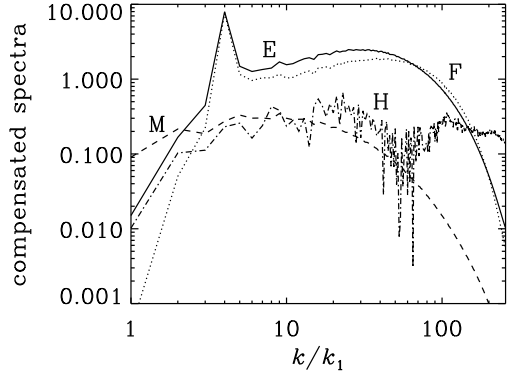}\end{center}\caption[]{
Compensated spectra of kinetic and magnetic energies and helicities
in the kinematic regime for $\Pm=1$.
The spectra are denoted by letters E, F, M, and H, as described in the text.
}\label{pspec_aver_comp_lopm_hel}\end{figure}

In order to judge the correspondence with various power-law scalings
we label, in \Fig{pspec_aver_comp_lopm_hel}, various compensated spectra
as follows:
\EQ
\mbox{label E}:\quad E(k)\epsilon_K^{-2/3}k^{5/3},
\EN
\EQ
\mbox{label F}:\quad |F(k)|\epsilon_K^{-2/3}k^{5/3}/2\kf,
\EN
\EQ
\mbox{label M}:\quad M(k)\kf(k/k_*)^{-3/2}/M,
\EN
\EQ
\mbox{label H}:\quad |H(k)|\kf\epsilon_K^{-2/3}k^{5/3} E/2M,
\EN
where $k_*=\int k M(k)\,\dd k/M$ is the wavenumber
where the magnetic energy spectrum peaks and $\epsilon_K$ is the
kinetic energy dissipation per unit mass.
The compensated kinetic energy spectrum shows a bottleneck that is clearly
stronger than in the case without helicity (e.g., Kaneda et al.\ 2003,
Haugen \& Brandenburg 2006).
The kinetic helicity spectrum shows a similar spectrum that also has a
strong bottleneck, which is particularly evident when it is compensated
by $k^{5/3}$.
The existence of a $k^{5/3}$ subrange for the modulus of the kinetic helicity
spectrum is well known from early closure calculations
(Andr\'e \& Lesieur 1977), and has also been seen in direct numerical
simulations (Borue \& Orszag 1997, Brandenburg \& Subramanian 2005a)
and in shell model calculations (Ditlevsen \& Giuliani 2001).
Such a scaling implies that the relative spectral kinetic helicity,
\EQ
{\cal R}_K(k)\equiv F(k)/2k E(k),
\EN
decreases toward small scales like $k^{-1}$
and has a maximum at $k=\kf$ with ${\cal R}_K(\kf)=0.96$.
At that scale, the relative magnetic helicity,
\EQ
{\cal R}_M(k)\equiv kH(k)/2 M(k),
\EN
is $-0.15$, $-0.08$, and $+0.42$ for $\Pm=1$, 0.1, and 0.01, respectively.
The realizability condition implies that the moduli of ${\cal R}_K(k)$
and ${\cal R}_M(k)$ are less than unity (Moffatt 1969).
The positive sign for $\Pm=0.01$ agrees with the idea that the helical
driving of the flow imprints a helical field of the same sense at the
same scale.
Owing to an inverse cascade of magnetic helicity (Pouquet et al.\ 1976),
$H(k)$ is of opposite sign at large scales.
While this is very clearly established in the nonlinear regime (B01)
or for small values of $\Pm$, a larger range of scales attains negative
values during the linear stage when $\Pm=0.1$ and 1.

\begin{figure}[t!]
\begin{center}\plotone{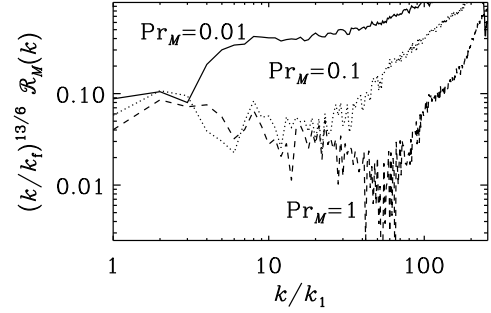}\end{center}\caption[]{
Relative spectral magnetic helicity, $k|H(k)|/2M(k)$, compensated
by $(k/\kf)^{13/6}$, for $\Pm$ ranging from 0.01 to 1.
}\label{pspec_aver_comp_lopm_hel3}\end{figure}

For the magnetic helicity we also find an approximate $|H(k)|\sim k^{-5/3}$
spectrum, which is different from the nonlinear case when the current helicity,
$C(k)=k^2H(k)$ shows a $k^{-5/3}$ spectrum (Brandenburg \& Subramanian 2005a).
Assuming that $M(k)\sim k^{3/2}$, the relative spectral helicity
would seem to decrease now more rapidly like $k|H(k)|/2M(k)\sim k^{-13/6}$.
\FFig{pspec_aver_comp_lopm_hel3} shows that the correspondingly compensated
magnetic helicity to energy ratio changes now less strongly in the range
$4\leq k/k_1\leq40$.

\subsection{Saturation regime}

\begin{figure}[t!]
\centering\includegraphics[width=\columnwidth]{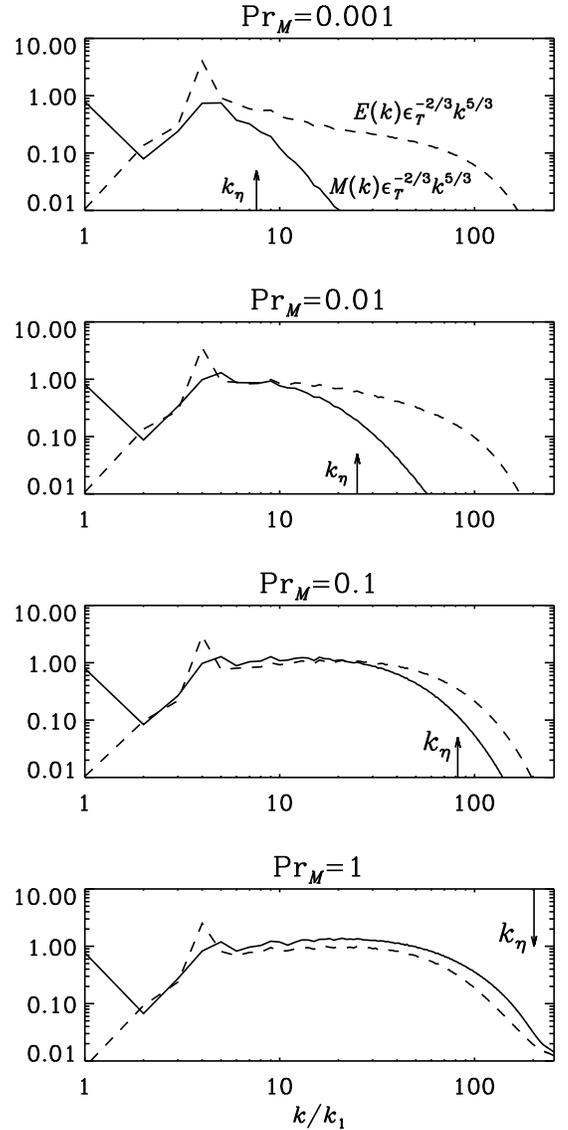}\caption{
Kinetic and magnetic energy spectra in the saturated regime for
$\Pm=10^{-3}$ with $\Rey=4400$, $\Pm=10^{-2}$ with $\Rey=2300$,
$\Pm=0.1$ with $\Rey=1200$, and $\Pm=1$ with $\Rey=450$.
All spectra are compensated by $\epsilon_T^{-2/3}k^{5/3}$.
The ohmic dissipation wavenumber, $k_\eta=(\epsilon_M/\eta^3)^{1/4}$,
is indicated by an arrow.
The viscous dissipation wavenumbers are 430, 350, 290, and 180 for
$\Pm=10^{-3}$, $10^{-2}$, 0.1, and 1, respectively.
}\label{pspec_aver_comp_saturated2}\end{figure}

Eventually the initial exponential growth comes to a halt and is followed
by a resistively long saturation phase during which a large-scale magnetic
field develops at wavenumber $k_1$, regardless of the value of $\kf$.
Owing to the use of periodic boundary conditions,
this large-scale field tends to be force-free and fully helical, and its
energy per unit volume is by a factor $\kf/k_1=4$ larger than the value at
$\kf$, which in turn is comparable to the kinetic energy per unit volume.
For details see B01.
Here we only consider the end of this slow saturation phase.
Compensated kinetic and magnetic energy spectra are shown in
\Fig{pspec_aver_comp_saturated2} for magnetic Prandtl numbers ranging
from 1 to down to $10^{-3}$.

In the final saturated state, and especially for $\Pm=1$, the $M(k)$
and $E(k)$ spectra are nearly on top of each other with $M(k)$ being
slightly larger than $E(k)$ by 20\%, which is qualitatively similar to
the non-helical case (cf.\ Haugen et al.\ 2003).
There are indications of a somewhat shallower spectrum due to a bottleneck
effect both for kinetic and magnetic energies just before the two enter
the viscous and resistive dissipation ranges.
Also for $\Pm=0.1$ there is a short range where $M(k)$ exceeds $E(k)$,
but then, not surprisingly, $M(k)$ turns into the dissipation range
before $E(k)$ does.
This is even more clearly the case for $\Pm=0.01$.

Low-$\Pm$ turbulence has the interesting property that for given
numerical resolution much larger fluid Reynolds numbers can be achieved
than for $\Pm=1$.
This is simply because almost all the energy is dissipated resistively,
and the energy that continues along the kinetic energy cascade is
comparatively weak, so not much viscosity is needed for dissipating
the remaining kinetic energy.
In fact, for $\Pm=0.01$ we were able to go to $\Rey=2300$ with a resolution
of only $512^3$ meshpoints.
For $\Pm=0.1$ and 1 and the same resolution we could only go to
$\Rey=1200$ and 450, respectively.

\subsection{Diverting most of the energy into Joule heat}

\begin{figure}[t!]
\begin{center}\plotone{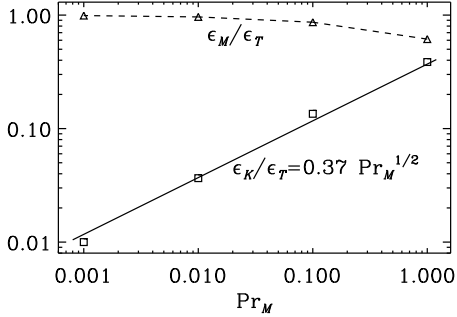}\end{center}\caption[]{
Dependence of the fractional kinetic and magnetic energy dissipation rates.
Note that the fractional kinetic energy dissipation decreases with
decreasing $\Pm$ to the 1/2 power.
}\label{pepsKM}\end{figure}

As has recently been stressed by Mininni (2007), an increasing fraction of
energy is being dissipated via Joule dissipation, as $\Pm$ decreases,
In \Fig{pepsKM}, we plot the dependence of the kinetic and magnetic
energy dissipation rates per unit mass,
$\epsilon_K=\bra{2\rho\nu\SSSS^2}/\rho_0$ and
$\epsilon_M=\bra{\eta\mu_0\JJ^2}/\rho_0$, relative to the total dissipation,
$\epsilon_T=\epsilon_K+\epsilon_M$, versus $\Pm$.
The data are well described by a power-law fit of the form
\EQ
\epsilon_K/\epsilon_T\approx0.37\,\Pm^{1/2}.
\EN
Thus, for $\Pm=1$ about the 37\% of the energy is dissipated into
viscous heat, while 63\% is dissipated via Joule dissipation.
This is similar to the case of non-helical hydromagnetic turbulence
(Haugen et al.\ 2003), where these numbers are about 30\% and 70\%,
respectively.

In turbulence the energy dissipation is generally proportional to $U^3/L$,
where $U$ is the typical velocity and $L$ is a typical length scale.
Conventionally one defines a dimensionless dissipation parameter as
\EQ
C_\epsilon={\epsilon_T\over U^3/L},
\EN
where $U$ is the one-dimensional rms velocity, which is related to
$\urms$ via $U^2=\urms^2/3$, and $L$ is the integral scale and is
related to $\kf$ via ${3\over4}\pi/\kf$.
In non-helical turbulence this value is typically around 0.5
(see also Pearson et al.\ 2004), but this value has never been
determined for hydromagnetic turbulence with helicity.
An exception is the work of Blackman \& Field (2008), who considered a range
of power-law scalings for kinetic and magnetic energy spectra to calculate
analytically the dissipation rates.

\begin{figure}[t!]
\begin{center}\plotone{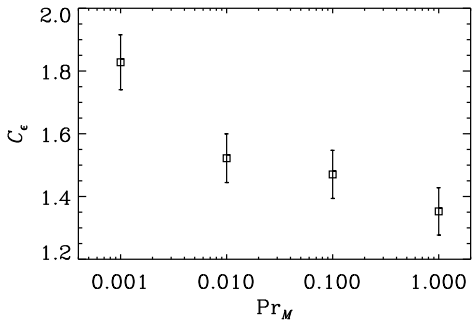}\end{center}\caption[]{
Dimensionless total energy dissipation rate, $C_\epsilon$,
as a function of $\Pm$.
Error bars have been estimated based on averages taken over
each third of the full time series.
}\label{pepsT}\end{figure}

It turns out that for our runs, $C_\epsilon\approx1.5$, i.e.\
$\approx3$ times larger than the usual value; see \Fig{pepsT}.
Let us now discuss possible reasons for this difference.
In the definition of the quantity $C_\epsilon$ one assumes that
the energy flux scales with $U^3/L$.
However, $U$ is based on the typical rms velocity.
In the presence of a strong dynamo-generated magnetic
field it may be sensible to base it on a combination
of typical velocity and magnetic field strength.
In our case we have $\bra{\BB^2/\mu_0}/\bra{\rho\UU^2}\approx2$, so
$U$ would need to be scaled up by a factor $\sqrt{3}$, which reduces
$C_\epsilon$ by a factor $3^{3/2}\approx5$ to about 0.3.
This value is nearly independent of the value of $\Pm$, which was
also found by Blackman \& Field (2008) under plausible assumptions.

\begin{figure}[t!]
\centering\includegraphics[width=\columnwidth]{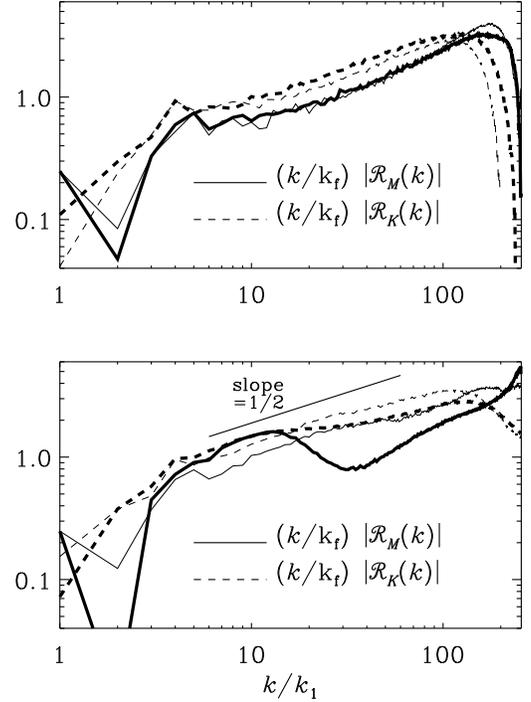}\caption{
Spectral current and kinetic helicity ratios for the same four
runs shown in \Fig{pspec_aver_comp_saturated2}.
The upper panel is for $\Pm=1$ (thin line) and 0.1 (thick line), while
the lower panel is for $10^{-2}$ (thin line) and $10^{-3}$ (thick line).
Note that for $\Pm=1$ and 0.1 the profiles of
$(k/\kf)|{\cal R}_M(k)|$ and $(k/\kf)|{\cal R}_K(k)|$
are reasonably flat in the range $6\leq k/k_1\leq14$.
The 1/2 slope is shown for comparison.
}\label{pspec_aver_comp_saturated3}\end{figure}

\begin{figure*}[t!]
\begin{center}\plotone{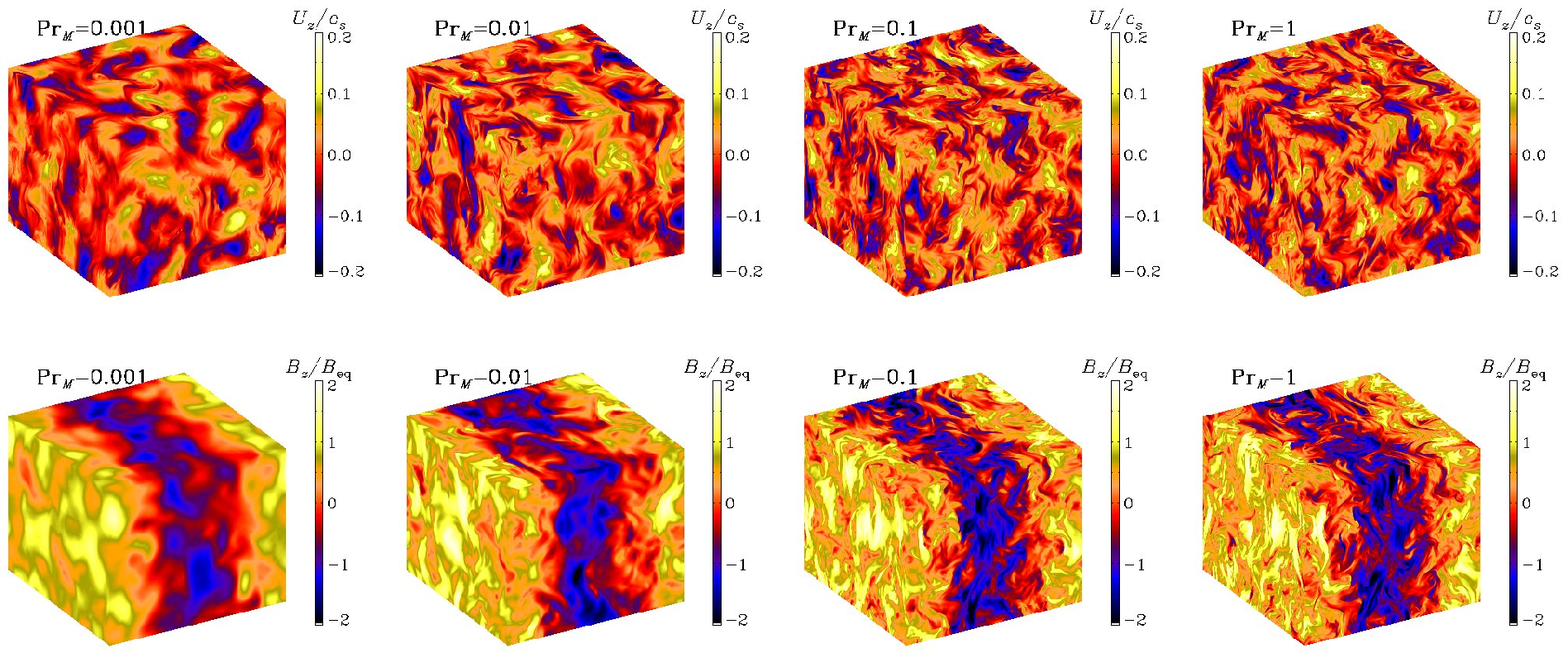}\end{center}\caption[]{
Visualizations of $B_z$ and $U_z$ for $\Pm=10^{-3}$ at $\Rey=4400$
(left), $\Pm=10^{-2}$ at $\Rey=2300$, $\Pm=0.1$ at $\Rey=1200$, and
$\Pm=1$ at $\Rey=450$ (right).
The orientation of the axes is the same as in \Fig{B}.
}\label{UB}\end{figure*}

\subsection{Helicity spectra}

As mentioned before, in the nonlinear regime both kinetic and current
helicities, $F(k)$ and $C(k)=k^2H(k)$, respectively, are expected
to display a forward cascade with a $k^{-5/3}$ spectrum.
If this is true, we would expect that within some wavenumber interval
$|{\cal R}_K(k)|$ and $|{\cal R}_M(k)|$ decrease with increasing $k$
like $k^{-1}$.
In \Fig{pspec_aver_comp_saturated3} we show the correspondingly compensated
relative kinetic and magnetic helicity spectra.
It turns out that they are surprisingly similar regardless
of the value of $\Pm$.
For $\Pm=1$ and 0.1 the compensated profiles of ${\cal R}_M(k)$ and
${\cal R}_K(k)$ are reasonably flat in the range $6\leq k/k_1\leq14$.
However, for $\Pm=10^{-2}$ and $10^{-3}$ the compensated profiles show
an increase proportional to $k^{1/2}$.
The fact that the anticipated $k^{-1}$ scaling occurs only for magnetic
Prandtl numbers down to 0.1 and only over an extremely short range
may indicate that our Reynolds numbers are still too small to yield
conclusive results.
Especially at smaller scales, and certainly in the runs with the smallest
$\Pm$, the compensated relative kinetic and magnetic helicity
spectra are compatible with a $k^{1/2}$ slope.
This would imply a $k^{-7/6}$ spectrum for the kinetic and current
helicities, which is shallower than that anticipated  for a forward
cascade, but still steeper than that in the case of equipartition.
We emphasize again that this applies to the resistively controlled regime.

In \Fig{pspec_aver_comp_saturated3} we see that at $k=\kf$ both
${\cal R}_K$ and ${\cal R}_M$ are close to unity.
This indicates that velocity and magnetic fields are nearly fully helical.
However, for $k>\kf$ the velocity and magnetic fields become less helical,
because the compensated relative helicities in \Fig{pspec_aver_comp_saturated3}
increase with $k$ not faster than to the 1/2 power.
On the other hand, for $k=k_1$ the magnetic field is again fully helical,
i.e.\ $(k_1/\kf)|{\cal R}_M(k_1)|$ is equal to $k_1/\kf=1/4$,
but its helicity has the opposite sign.

In \Tab{Trelhel} we compare the values of ${\cal R}_M(k)$ during the
linear and nonlinear stages at the wavenumbers $k_1$ and $\kf$
for the three or four values of $\Pm$.
Note that during the nonlinear stage ${\cal R}_M(k_1)$ and
${\cal R}_M(\kf)$ are of opposite sign.
The former is close to $-1$ while the latter increases from 0.52 to 0.72
as $\Pm$ decreases.
As already indicated in \Sec{Spectra}, during the linear stage, the two
are of opposite sign only for $\Pm=0.01$, while for larger values of $\Pm$
a larger range of scales appears to be affected by the inverse transfer
of magnetic helicity causing ${\cal R}_M(\kf)$ to be negative.
It would be tempting to try and model this behavior using, for example,
the four-scale helical dynamo model of Blackman (2003).

\begin{table}[htb]\caption{
Comparison of ${\cal R}_M(k_1)$ and ${\cal R}_M(\kf)$ during the linear
and nonlinear stages for different values of $\Pm$.
}\vspace{12pt}\centerline{\begin{tabular}{ccccc}
\hline
\hline
& \multicolumn{2}{c}{linear} & \multicolumn{2}{c}{nonlinear} \\
$\quad\Pm\quad$
& $\quad k_1\quad$ & $\quad\kf\quad$
& $\quad k_1\quad$ & $\quad\kf\quad$ \\
\hline
$10^{-3}$ &         &         & $-0.993$ & $+0.72$ \\
$10^{-2}$ & $-0.88$ & $+0.42$ & $-0.994$ & $+0.66$ \\
$10^{-1}$ & $-0.59$ & $-0.08$ & $-0.993$ & $+0.59$ \\
1         & $-0.41$ & $-0.15$ & $-0.993$ & $+0.52$ \\
\hline
\label{Trelhel}\end{tabular}}\end{table}

\subsection{Effects on the velocity pattern}
\label{VelocityPattern}

In \Fig{UB} we compare visualizations of $B_z$ and $U_z$ for all four
values of $\Pm$.
The velocity and magnetic field patterns are surprisingly similar
for all four values of $\Pm$.
Only for $\Pm=10^{-2}$ and $10^{-3}$ the magnetic field appears noticeably
smoother than in the other two cases.
The velocity field shows a marked anisotropy with small-scale elongated
patterns aligned with the local direction of the mean magnetic field, which
is here of the form $\meanBB\sim(0,\sin k_1x,-\cos k_1x)$.
The anisotropy in the velocity can still be seen for small values of
$\Pm$, but the small-scale patterns are slightly smoother.

\section{Conclusions}

In many astrophysical bodies the magnetic Prandtl number is small,
while in most simulations its value is chosen to be close to unity.
As we have shown here, this mismatch is of relatively minor consequence
for large-scale dynamos that are driven by helical forcing.

In the nonlinear stage, the velocity and magnetic field patterns are
remarkably independent of the value of $\Pm$.
The only thing that changes is the length of the inertial range.
A small magnetic Prandtl number simply means that the
magnetic energy spectrum turns into the dissipation range more quickly
than the kinetic energy spectrum.
It also means that essentially all the energy is dissipated via Joule heat.
This was recently also demonstrated by Mininni (2007).
One reason is that the case of fully helical turbulence
studied in the present paper is a particu\-larly
simple one, because it leads to uniform mean-field dynamo action
with large-scale pattern formation covering the entire domain.
In this paper we have seen that, at least for values of $\Rey$ up to
4400, the dynamics of this large-scale pattern, i.e., of the large-scale
magnetic field, is quite independent of how long the inertial range of
the turbulence is.
In the absence of helicity, there is only small-scale dynamo action,
which is driven by the dynamics at the smallest possible scale,
i.e.\ the resistive scale.
In that case it does matter what the dynamics of the turbulence is at
that scale.
However, in that case it has not yet been possible to find dynamo
action for values of $\Pm$ down to the values considered here.
Nevertheless, it is possible that even in that case there is an
asymptotic regime for large enough values of $\Rm$ where the dynamics
of the magnetic field is independent of the value of $\Pm$, even though
this regime is not yet accessible with present day computers.

To estimate the value of the magnetic Prandtl number in dense astrophysical
bodies, one has to use the {\it Spitzer} formulae for $\eta$ and $\nu$.
The resulting magnetic Prandtl number is (e.g., Brandenburg \& Subramanian 2005b)
\EQ
\Pm=1.1 \times10^{-4}
\left({T\over10^6\K}\right)^4
\left({\rho\over0.1\g\cm^{-3}}\right)^{-1}
\left({\ln\Lambda\over20}\right)^{-2},
\EN
so at the bottom of the solar convection zone the magnetic Prandtl number
is clearly rather small ($\sim10^{-4}$).
Nevertheless, simulations of solar and stellar dynamos available so far
$\Pm$ are set to values of the order or unity.
Although we have shown here that the resulting large-scale fields are
similar to the more realistic case of small values of $\Pm$,
an important difference is that the small-scale dynamo may be more
pronounced when $\Pm$ is of order unity.
In practice this means that a positive detection of dynamo action
in a simulation might not necessarily be relevant for understanding
the Sun, unless suitable conditions for the excitation of large-scale
dynamo action are also met.
On the other hand, once the large-scale dynamo is really excited,
and if it is fully saturated, it is then quite feasible to lower the
value of $\Pm$ significantly---without losing the large-scale dynamo.
In fact, lowering $\Pm$ in a saturated large-scale dynamo means that
most of the energy will be dissipated via Joule heating, and that the
kinetic energy cascade only carries a small fraction of the total energy.
This allows us to  increase the value of $\Rey$, and hence to decrease
the viscosity and thereby the value of $\Pm$ even further.
Simulations of large-scale dynamo action in turbulent convection
(K\"apyl\"a et al.\ 2008) provide one example where it is indeed feasible
to lower $\Pm$, although in that case the system is not uniform and so
energy dissipation via Joule heating is only possible in those locations
where the dynamo is strong enough (P.\ J. K\"apyl\"a 2008, private communication).

\acknowledgements
I thank Eric G.\ Blackman and Pablo Mininni for useful comments on the
paper, and an anonymous referee for spotting a number of errors in the
original version.
It is a pleasure to acknowledge the organizers of the KITP program
on dynamo theory and the staff of the KITP for providing a stimulating
atmosphere.
This research was supported in part by the National Science Foundation
under grant PHY05-51164 and the Swedish Research Council under grant
621-2007-4064.
The computations have been carried out at the National Supercomputer
Centre in Link\"oping and at the Center for Parallel Computers at the
Royal Institute of Technology in Sweden.


\end{document}